# On the unusual behavior of nitride compounds

W. T. Zheng and Chang Q. Sun[*]

*Department of Materials Science, Jilin University, Changchun 130012, P.R. China, E-mail: wtzheng@jlu.edu.cn*

*School of Electrical and Electronic Engineering, Nanyang Technological University, Singapore 639798*

*Abstract*

This report presents consistent insight into the mechanism behind the unusual behavior of nitride compounds from the perspective of tetrahedron bond formation and its consequence on valence density of states. An extension of the recent bond-band-barrier (BBB) correlation mechanism for oxidation [Sun CQ, *Prog Mater Sci* 2003;48:521-685] to the electronic process of nitridation has led to the essentiality of sp-orbital hybridization for a nitrogen atom upon interacting with atoms in solid phase of arbitrary less-electronegative element. In the process of nitridation, a nitrogen atom forms a quasi-tetrahedron with surrounding host atoms through bonding and nonbonding interaction associated with production of electronic holes and antibonding dipoles, which add corresponding density of states to the valence band of the host. It is suggested that the valance alteration of the system takes the responsibility for the blue shift in photoluminescence, lowered work function for cold cathode field emission, corrosion and wear resistant, high elasticity, and magnetic modulation as well.

---

[*] Fax: 65 6792 0415; *E-mail:* ecqsun@ntu.edu.sg; URL: www.ntu.edu.sg/home/ecqsun/

*I       Introduction*

Nitride compounds have formed a class of materials with fascinating properties that have widely been used for mechanical and elastic enhancement, wear and corrosion resistant, photon and electron emission, as well as magnetic modulation [1]. For instances, investigating the Fe-nitride thin films [2,3,4,5] uncovered that the crystal structures, saturation magnetization ($M_S$) and Curie temperature ($T_C$) of the Fe films could be modulated by adjusting the concentration of $N_2$ in the mixture of Ar and $N_2$ sputtering gases. The $M_S$ value of $\alpha''$-$Fe_{16}N_2$ phase is ~25% higher than that of the pure Fe (2.22 $\mu_B$) and then the $M_S$ value drops with increasing the N concentration in the films with crystal structure variation from $\varepsilon$-, $\gamma$-, $\xi$-, to amorphous phase. Amorphous FeN films exhibit however paramagnetic features. An addition of N to the rare earth (R)-ferromagnetic (Co and Fe) system increases the $M_S$ value and raises the $T_C$ considerably [6,7]. The $M_S$ of the N-R(Fe, Co) alloys were increased by about 30 ~ 40 % relative to their parent alloys [7]. The N-modulated $M_S$ may enable these kinds of materials to be used in high-density data storage and used as new kinds of strong permanent magnets. Inclusion of N in the synthetic diamond films could significantly reduce the threshold of cold cathode emission of diamond thin films [8]. The threshold is even lower than those doped with boron and phosphorous. The work function of carbon nitride films could be reduced to ~0.1 eV deposited at 200 $^o$C substrate temperature under 0.3 Pa nitrogen pressure [9] in sputtering. Nitridation can turn a conductor into a semiconductor or even an insulator, such as AlN [10], GaN [11], and InN [12]. N-based group-III and -IV semiconductors are prosperous materials for blue and green light emitting [13]. These wide-band nitride semiconductors have been commercially available for applications in flat-panel displays and blue-ultraviolet laser diodes that promise high-density optical data storage, optic-communication and high-resolution laser printing. Although experimental indications were unsatisfactory in the search for a super-hard phase of carbon nitride [14,15,16,17,18,19,20,21,22], the nitride films shown surprisingly high elasticity (~100%) and high mechanical strength at relatively low indentation load (< 1 mN) [21,23,24,25,26].

Even though the performance of nitride compounds has been intensively investigated and widely used, understanding the correlation between the local atomic-bonding or the energy-band structures and the observed, or predicted properties of such compounds is still infancy [21,23,27]. Concerns about the correlation between the chemical bond and valence density of states and their effects on materials performance are therefore necessary for deeper and consistent insight into the nitride systems towards predictable design of functional materials. In the present report, we extend the recent bond-band-barrier (BBB) correlation mechanism originated for oxidation [28] to the unusual performance of a nitride towards consistent insight into the chemical stability, mechanical strength, magnetic tunability and electronic and optical emission properties.

*II       Principle: bond-band-barrier (BBB) correlation*
2.1     Nitride tetrahedron formation

Figure 1a illustrates the physical model of a nitride ($NA_4$) quasi-tetrahedron obtained by replacing the hydrogen in the $NH_3$ molecule with atoms of arbitrary element *A* that should be less electronegative than N. Upon interacting with atoms *A* in a solid phase, the sp-orbital of the N atom hybridizes essentially into four directional orbitals. To fill up the four directional sp-hybridized orbitals of the N atom which has five $2s^22p^3$ valence electrons, three more electrons are required from the A neighbors. Therefore, among the four hybridized orbitals, three are occupied by bonding electron pairs shared by the N and the A atoms and one orbital is occupied by the nonbonding electron lone pair of the N. Because of its higher electronegativity (3.0 without unit), the N atom acts as electron acceptor rather than a donor and therefore the three bonds are mainly polar-covalent. The nonbonding lone pair formation is an intrinsic property of nitrogen and oxygen upon the sp-orbital hybridization, which is independent of whatever the constituent *A* is. However, the lone pair tends to polarize the atom (dipole formation in Figure 1) on which the lone pair is acting. In the nitride tetrahedron, four A atoms of different valences surround the central N atom. The smaller $A^+$ ions donate electrons to the N acceptor, $N^{3-}$. The atoms labeled 2 are the lone-pair-induced dipoles, $A^{dipole}$. As consequences of tetrahedron bond formation, the positive ions also polarize their neighbors to form antibonding dipole states as well. As noted by Atkins [29], the antibonding states are dominated by the less electronegative elements, and therefore, antibonding interaction forms between the $A^{dipole} \leftrightarrow A^{dipole}$ dipoles rather than between the $N^{3-}$ and the $A^+$ ions. In a particular nitride system, the geometrical environment determines the orientation of the nitride tetrahedron. The difference between a nitride tetrahedron and an oxide tetrahedron (Figure 1b)[27] is that one oxide tetrahedron has two lone pairs while the nitride has one, which follows a '4-n' rule, where is the valence value. This difference determines that the group symmetry of a nitride tetrahedron differs from that of an oxide tetrahedron. Such a small difference determines that a nitride performs entirely different from an oxide. If the acceptor is over dosed, a hydrogen-like bond is formed, which transports electrons from the antibonding states to the bonding orbitals of the acceptors [27].

Figure 1 Primary (a) $NA_4$ nitride and (b) $OA_4$ quasi-tetrahedron model [27]. Smaller ions donate electrons to the central N/O acceptor of which the sp-orbital hybridizes with production of a nonbonding lone pair. The atoms labeled 2 are the lone-pair-induced A dipole. O/N hybridizes and interacts with arbitrary element A through bonding and nonbonding lone-pair to form the quasi-tetrahedron: $NA_4 = N^{3-} + 3A^+$ (labeled 1) + $A^{dipoe}$ (labeled 2), and $OA_4 = O^{2-} + 2A^+$ (labeled 1) + $2A^{dipoe}$ (labeled 2). The number and orientation of the lone-pair give different geometrical symmetry as indicated.

2.2     Valence density of states (DOS)

Because of the $NA_4$ and $OA_4$ tetrahedron formation, the energy band of the A host solid is modified with four additional DOS features. As illustrated in Figure 2, the difference of the DOS between a nitride or oxide compound and its parent solid leads to the bonding ($< E_F$), nonbonding ($\leq E_F$), holes ($\leq E_F$) and antibonding ($> E_F$) states [27,28]. The positions of the electron holes in energy space are located at the valence-band upper edge of a semiconductor or below the $E_F$ of a metal. The bonding states are slightly lower while the nonbonding states are around the original *p*-levels of an isolated N/O atom, as nonbonding states neither add nor decrease the system energy [29]. Electrons of the dipoles gain energy due to the polarization and then occupy the empty states well above $E_F$. The bonding and antibonding processes yield holes under $E_F$, which turns a metal to be a semiconductor such as group-III nitride/oxide compounds [13]. Due to the hole-production, the effective band-gap of an intrinsic semiconductor is chemically widened such as the cases of Ge and Si nitride/oxide [30]. The dipole formation enhances the charge density of the surface atomic layers and hence reduces the work function of the system [31]. The alteration of atomic sizes and valances modifies the potential barrier or the morphology of the surface.

Figure 2 N and O induced valence DOS differences between nitride/oxide compounds and the parent metal (upper) or the parent semiconductor (lower). The N or O modifies the energy-band of the host by adding four features through the following processes [27]:

(i) Charge transport from A to the accepter to form the $sp^3$ – hybrid bonding, which produces states at locations slightly lower than the 2p-level of the acceptor;

(ii) The $sp^3$ – hybridization produces nonbonding lone pair states that neither raise nor lower the system energy;

(iii) Lone pairs and ions induce antibonding dipoles that form states well above the $E_F$, which lowers the work function; and

(iv) Formation of bonding and anti-bonding generates electron-hole states close to $E_F$ of a metal or near the valence band edge of a semiconductor, which changes a conductor to be a semiconductor or widens the band gap of a semiconductor.

(v) Additional accepter will attract the electrons of the dipoles to form Hydrogen-like bond that narrows the antibonding density of states and restores the work function.

*III  Verification*
3.1  Atomic valence and bond geometry

The $C_{3v}$ symmetry of a $NA_4$ cluster can be evidenced directly by the fact that most of the nitrides prefer the hcp(0001) or the fcc(111) orientation, such as AlN, GaN, TiN, etc. Figure 3 shows the scanning electron microscopy images of SiCN crystals grown using microwave-assisted CVD on Si substrate with the gas mixture of $N_2$ + $CH_4$ [32]. It is apparent that the SiCN crystallites prefer the hcp columnar structures.

Figure 3 SEM images of the SiCN crystals formed on Si substrate with a 10/4 sccm $N_2/CH_4$ gas mixture for (a) 4 hrs and (b) 10 hrs and (c) 50/1 sccm for 10 hrs. SiCN crystallites prefer the hcp structures [32].

N-chemisorption study by Sotto *et al.* [33] suggested that the $Cu_3N$ forms on the Cu(001) and Cu(h11) surfaces with nitride patches that are imaged with the scanning tunneling microscopy (STM) as being ~ 0.8 Å below the clean surface. The STM depression often corresponds to ions or atomic vacancy. Low-energy-electron diffraction (LEED) study of the N-Ru(0001) surface [34] reveals that the N sinks deeply into the threefold hollow-site of the top layer with radius-away reconstruction, indicating a central position of the N in the $NRu_4$ cluster of $C_{3v}$ symmetry with lone pair directing downward the surface according to the current premise of tetrahedron formation. However, for N-Ni(001) surface, the lone pair is directed sideway into the open end of the surface and the electrostatic interaction between the alternative $Ni^+$ and $Ni^{dipole}$ drives the Ni(001) surface to be reconstructed with ohombi-chain forming along the <11> directions [35] (see Figure 4). Compared with C adsorption that produces compressive stress, N addition leads to tensile stress at the surface due to the different atomic valences. Understanding of the N and C induced Ni(001) surface reaction not only gives a consistent insight into the mechanism for the clock-and-anticlock wise reconstruction but also a novel approach of TiCN graded buffer layer to neutralize the interfacial bond stress and hence enhance the diamond–metal adhesion substantially [36].

Figure 4 (a) STM image and (b) the corresponding bond network for N-induced Ni(001) surface reconstruction with rhombus-chain formation along the <11> direction. The atoms labeled 1 and 2 in the basic tetrahedron represents $Ni^+$ and $Ni^{dipole}$ [35]. The depressions link two $Ni^+$ at the surface.

## 3.2 Valence DOS formation and lone-pair vibration

Compared with the nanometric SiC, nanostructured SiN exhibits two extra DOS features [37]. One is located at ~3.3 eV below $E_F$ and the other is 1 ~ 3.8 eV above $E_F$. The feature below $E_F$ was identified as the N-2p lone-pair π-orbital at the upper edge of the valence band of SiN. The latter unknown feature can readily be ascribed as the contribution from the lone-pair induced antibonding states according to the current BBB correlation. A first-principle calculation [17] predicted that the N-N lone-pair repulsion in the carbon nitride leads to a ~ 2.2 eV elevation in anti-bond energy. The *ab initio* calculations of the N-Ru(0001) surface [34] and O-Ru($10\bar{1}0$) surface [38] reveals the similar DOS features at +3.0 (antibond), -1.0 (holes) -3.0 (nonbond) and -6.0 eV (bond) around $E_F$, as compared in Figure 5, which are consistent surprisingly well with the current BBB correlation premise (Figure 2) that correlates the calculated DOS to the individual atomic valence alteration. Figure 6 shows the evolution of the valence DOS features upon C being gradually replaced by N in TiC compound. The shaded broad peak around $E_F$ can be attributed to the lone pair DOS whereas the energy of the antibonding DOS is beyond the scope of normal x-ray photoelectron spectroscopy (XPS) but it is detectable with the inverse ultra-violet PS or scanning tunneling spectroscopy (STS) [28]. STS measurement [39] has revealed strong DOS features in the conduction band of $CN_x$ nanotubes near the Fermi level (-0.18 eV), evidencing the existence of the lone pair and dipole states. For carbide, neither lone pair nor anti-bonding dipole could form upon reaction. However, a ~ 2.31 eV antibond DOS has been obtained in calculating carbon nitride [17]. Table 1 compares the DOS features of metal and semiconductor nitrides supporting the model predictions.

The presence of the lone pair in both oxide and nitride has been further confirmed using the Raman spectroscopy showing vibration features below 1000 cm$^{-1}$ [23], as compared in Figure 7, which is consistent with the H bond vibration of bio-molecules such as protein and DNA [40]. Compared with the Raman spectra of nitrides, the lone pair features of nitrides are much stronger because of the different numbers of lone pair in a tetrahedron. It is unambiguous that N or O adds indeed an anti-bond sub-band above $E_F$ due to the dipole formation, disregarding the host element [27,28].

Figure 5 The tight-binding approximation of the difference in density of states, n(Ru+N) - n(Ru), between Ru(0001)-c(2×2)-N and Ru(0001) surface. Four features correspond to the antibonding (~3.0 eV), hole (-1.0 eV), nonbonding (-3.0 eV) and bonding (-6.0 eV) states [34].

Figure 6 XPS profiles show the evolution of the valence DOS from TiC to TiN showing the additional shaded features around $E_F$ [42] that can be attributed to the lone pair.

**Figure 7** Low-frequency Raman shifts indicate that weak bond interaction exists in (a) Zr (powder and sintered) and Al oxides and (b) Ti and amorphous carbon nitrides, which correspond to the nonbonding electron lone pairs generated during the sp-orbital hybridization. The number of lone pairs follows a "4-n" rule, where n is the valence value of the electronegative

additives. For oxides, nitrides and carbides, n = 2, 3 and 4, respectively. Therefore, the peak intensities of oxides are stronger than that of nitrides but there are no such peaks at all for amorphous carbon and tungsten carbide [23].

Table 1 N adsorbate-derived DOS features adding to the valence band of metals (unit in eV). Holes are produced below $E_F$. All the data were probed with XPS unless otherwise indicated.

| N-added surfaces | Anti-bond (dipole) > $E_F$ | Nonbond (Lone pair) < $E_F$ | Bond (sharing pair) < $E_F$ |
|---|---|---|---|
| N-Cu(001) [41] | 3.0 | -1.2 | -5.6 |
| N-Ru(0001) [34] | 3.0 | -3.0 | - 6.0 |
| TiCN [42] | | 0.0 ± 1.0 | -5.7 |
| a-CN [43] | | -4.5 | -7.1 |
| CN [44] | | -2.3 | |
| N-Ag(111) [45] | | -3.4 | -8.0 |

*III     Applications*
  3.1     Corrosion and wear resistivity

We may consider the case that the N reacts with a surface of $C_{3v}$ symmetry, such as fcc(111) and hcp(0001) planes (for **Figure 1**a example). The lone pair directs into the substrate and hence the surface is networked with the smaller $A^+$ and the fully bonded $N^{3-}$ ion cores with densely packed electrons. The outer shells of the $A^+$ ions are emptied due to charge transport from upper DOS states of the host to the lower empty DOS states of nitrogen upon bond formation. The reaction not only lowers but also densifies the DOS in the valence band due to $A^+$s and $N^{3-}$ formation. Therefore, the top surface layer should be inert in chemistry as it is harder for one additional accepter to catch electrons from the lower and denser DOS, or from the already bonded atoms compared with otherwise unbonded neutral atoms. This configuration may explain why a nitride surface is corrosion resistant.

The high intra-surface strength due to the ionic network is responsible for the hardness of the top layer. On the other hand, the $N^{3-}$ - $A^+$ network at the surface is connected to the substrate mainly through the nonbonding states. The nonbonding interaction is rather weak (~0.05 eV per bond) compared with the original metallic bonds (~1.0 eV per bond). The lone-pair weak interaction should be highly elastic, which makes the two adjacent layers more elastic at a pressing load lower than a critical value that breaks the weak interaction. The enhanced intra-layer strength makes a nitride usually harder (ultra hard ~20 GPa), and the weakened inter-layer bonding makes the nitride highly elastic and self-lubricative. Nanoindentation profiles from TiCrN surface and sliding friction measurements from CN and TiN surfaces have confirmed the predictions of high elasticity and high hardness at lower pressing load and the existence of the critical scratching load [23]. As compared in Figure 8 (a) and (b), the elastic recovery for GaAlN film [26] is as high as 100% under 0.7 mN indentation load compared with that of amorphous carbon film under the same pressing load. The GaAlN surface is also much harder than the amorphous-C film under the lower indentation load. Figure 8 (c) and (d) show the profiles of pin-on-disk sliding friction test, which revealed the abrupt increase of the friction coefficient under higher load [23]. For polycrystalline diamond thin films, no such abruption in friction coefficient is observed though the friction coefficient is generally higher than the carbide films. The absence of lone pairs in a-C film makes the film less elastic than a nitride under the same pressing load. The abruption in the friction coefficient means the existence of critical load that breaks the nitride interlayer bonding – lone pair interaction. For CN and TiN, the elastic recovery ranges from 65% to 85% with higher pressing load (5 mN) of indentation [21]. Therefore, the non-bonding interlayer interaction enhances the elasticity of nitride surfaces at pressing load lower than the critical values. Such high elasticity and high hardness by nature furnishes the nitride surfaces with self-lubricative for nano-tribological applications.

Figure 8 Mechanical strength and elasticity of nitride films. Comparison of (a) the 100% elastic recovery of GaAlN/Al$_2$O$_3$ surface at 0.7 mN load and (b) the 65% that of amorphous carbon in the same scale of indentation load. Pin-on-disk measurement of sliding friction shows the abrupt increase of the friction coefficient for (c) C nitride and (d) Ti nitride [23] evidences the critical load that breaks the surface bond. (e) The friction coefficient of diamond thin films show no abruption features though the coefficient is generally higher than the carbides.

3.2     Mechanical strength – harder than diamond?

The search for an exceptionally hard material of C-nitride has been lasted for more than one decade. The hardness and the elastic modulus of β-C$_3$N$_4$ are predicted to be comparable to or exceeding those of diamond but experimental data have shown no satisfactory though tremendous efforts have been made [46,47]. Theoretical investigations [16-19] modified the original C-N covalent-bond iteration, indicating that the superhard C$_3$N$_4$ phase can only be produced at higher pressure (68 Gpa). Taking the nonbonded N-N repulsion into consideration, the hypothetic stoichiometry C$_3$N$_4$ ratio has been suggested to be forbidden, and the N concentration should be no more than 50% [16,17]. Actually, experiments revealed that the hardness decreases with increasing N content. For instance, the elasticity of the CN has been measured to vary with substrate temperature and the N content. Increasing the substrate temperature from 100 to 350 °C at 2.5 mTorr N$_2$ pressure, the elastic recovery increases from ~ 60 to ~ 90% [23]. At substrate temperature 350 °C, N$_2$ pressure increases from 2.5 to 10 mTorr could reduce the elasticity to 68% [25]. Although the hardness of CN available to date [21] (~60 GPa) is below that of a diamond (~100 GPa), the elasticity has been confirmed rather high with a critical load for plastic deformation. Agreement between predictions and observations supports the nonbonding repulsion model [14,17]. Therefore, the presence of the lone-pair, or the fifth electron in nitrogen, should prohibit the carbon nitride from being harder than a diamond. However, it might be possible to obtain the superhard phase by removing the fifth electron of the N in synthesizing the CN under extremely high pressure or high temperature as suggested in Refs [16-19]. The hardness of nanocrystalline/amorphous composites such as nc-TiN/a-Si$_3$N$_4$, nc-TiN/a-Si$_3$N$_4$/ and nc-TiSi$_2$, nc-(Ti$_{1-x}$Al$_x$)N/a-Si$_3$N$_4$, nc-TiN/TiB$_2$, nc-TiN/BN, approaches that of diamond because of the interfacial mixing effect.[48,49] It has been found that the hardness and elasticity of nanometric TiN/CrN and TiN/NbN multi-layered thin films increase with reducing the structural wavelength (optimal at 7.0 nm). [50,51]

3.3     Magnetic modulation

Figure 9 shows the N content dependence of the $M_S$ of Fe films deposited using vacuum arc technique. The trend agrees with that deposited using facing target sputtering method [2].

Figure 9 Nitrogen content dependence of saturation magnetization $4\pi M_S$ of the Fe-N films [52].

According to the Ising approximation, the overall $M_S$ under zero external field is determined by:

$$H_{exchange} = \sum_{<i,j>} J_{ij} S_i \cdot S_j \propto \frac{S_i S_j}{r_{ij}} \cos\theta_{ij}$$

(1)

$S_i$ and $S_j$ is the magnetic momentum of individual atom and the $J_{ij}$ is the coefficient of exchange interaction between momentum i and j. $\theta_{ij}$ represents the angle between the $S_i$ and $S_j$ moment. There several factors controlling the $H_{exchange}$ and hence the overall magnetization. $\theta_{ij}$ varies with external field; the $J_{ij}$ varies with atomic distance; and $S_i$ and $S_j$ vary with atomic valences. Chemical reaction not only changes the separation between the atoms but also the electron distribution in the orbitals that modify the atomic valence and hence the $S_i$ and $S_j$. If the separation is too large, the system will be paramagnetic disregarding the $S_i$ and $S_j$ values. Interatomic distance not only varies with the extent of reaction but also a function of atomic coordination [53]. Here we may focus on the N-modified $S_i$ and $S_j$ values of a Fe nitride, as a sample. It can be seen from Table 2 that the total angular momentum increases when the Fe alters its atomic valence to $Fe^{n+}$ (n is an integer) or $Fe^{dipole}$. In the former, the Fe donates 3d electrons to the N acceptor. In the latter, the Fe 3d electrons are propelled by the Coulomb potential of the lone pair to an outer shell of itself, 4p or 4d. The $N^{3-}$ and its electrons do not contribute to the magnetization. The total momentum of the $Fe^{n+}$ varies from 2.0 to 3.0 $\mu_B$ and then drop to 2.0 $\mu_B$ when the valence changes from 1 to 4. The momentum for $Fe^{dipole}$ is 4.0 ($3d^5 4s^2 4p^1$) or even 5.0 $\mu_B$ ($3d^5 4s^2 4d^1$). The average momentum of an isolated tetrahedron ($N^{3-}$ + $3Fe^+$ + $Fe^{dipole}$) is then 2.875 or 3.125 $\mu_B$, being 25 ~ 40% higher than that of a pure Fe (2.22 $\mu_B$ as measured). For amorphous FeN, every Fe atom has four N neighbors and the Fe becomes $Fe^{3+/dipole}$ ideally. The low angular momentum of the $Fe^{3+/dipole}$ and the expanded lattice should take the responsibility for ferro-para magnetic transition [2]. The magnetization of a system not only varies with the angular momentum of individual atoms but also their exchange interaction that depends inversely on the atomic separation. These estimations agree with the findings in Ref. [2-4]. For rare earth, the 4f electrons combine less tightly than the 3d electrons of transition metal to their ion cores. Therefore, it is more likely for the 4f electrons to jump to higher energy shells giving the more pronounced increase of $M_S$ as found in Ref. [7]. Apparently, this mechanism accounts for the experimental observations [2-7] more reasonably than the assumption that the N donates electrons to the Fe atoms to increase the spin angular momentum. Nevertheless, it is forbidden for the Fe atom to capture electrons from the highly electronegative nitrogen.

Table 2 Variations of angular momentum (unit in $\mu_B$) of Fe with its atomic states.

| Valence state | Configuration | $S = \Sigma S_i$ | $L = \Sigma L_i$ | $J \ast = \Sigma(L \pm S)_i$ |
|---|---|---|---|---|
| Fe | $3d^6 4s^2$ | 2 | 0 (L-frozen) | 2 (2.22) |
| $Fe^+$ | $3d^5 4s^2$ | 2.5 | 0 | 2.5 |
| $Fe^{2+}$ | $3d^5 4s^1$ | 2.5 + 0.5 | 0 | 3.0 |
| $Fe^{3+}$ | $3d^4 4s^1$ | 2.0 + 0.5 | 0 | 2.5 |
| $Fe^{4+}$ | $3d^3 4s^1$ | 1.5 + 0.5 | 0 | 2.0 |
| $Fe^{dipole-1}$ | $3d^5 4s^2 4p^1$ | 2.5 + 0.5 | 0 + 1 | 4.0 |
| $Fe^{dipole-2}$ | $3d^5 4s^2 4d^1$ | 2.5 + 0.5 | 0 + 2 | 5.0 |

\* The total angular momentum is governed by the Hund's rule. Fe$^{\text{dipole-2}}$ corresponds to the antibonding states being well above the E$_F$.

### 3.4    Work function reduction for field emission

The work function ($\Phi$) or the threshold ($V_T$) in cold-cathode field emission of materials such as diamond, diamond like carbon (a-C) or carbon nanotubes (CNTs), can be modulated by doping proper amount of properly selected elements besides the geometric enhancement of the emitters. An addition of N to the CVD polycrystalline diamond thin films significantly reduces the $V_T$ of the diamond. The $V_T$ of the N-doped diamond is even lower than the $V_T$ of the diamond doped with boron and phosphorous [8,54,55,56,57], as compared in Figure 10.

Figure 10 N, P and B dipping effect on the threshold of cold field electron emission of diamond [8].

Zheng et al. [7] deposited carbon nitride films by rf reactive magnetron sputtered carbon in an $N_2$ discharge. Figure 11 compares the effect of processing parameters, such as nitrogen partial pressure, substrate temperature, and substrate bias on the field emission properties. The effective work function for deposited carbon nitride films determined using the Fowler-Nordheim equation is in the range of 0.01-0.1 eV at 200 $^{\circ}$C substrate temperature under 0.3 Pa nitrogen partial pressure [9]. Li et al. [58] examined the effect of nitrogen-implantation on electron field emission properties of amorphous carbon films. From the Fowler-Nordheim plots, they found the threshold field is lowered from 14 to 4 V/μm with increasing the dose of implantation from 0 to $5\times10^{17}$ cm$^{-2}$ and the corresponding effective work function is estimated to be in the range of 0.01-0.1 eV, as illustrated in Figure 11d. Boron nitride coated graphite nanofibers also emit electrons at much reduced $V_T$ (from 1.5 to 0.8 V/μm) with high (10$^2$ level) current intensity compared with the uncoated carbon fibers [59]. It is explained that introducing BN nanofilm to the surface leads to a significant reduction in the effective potential barrier height, or the $\Phi$. A tendency of N-buckling outward the BN nanotubes has been derived theoretically, which was explained as arising from the different hybridizations of B and N in the curved hexagonal layer and the N-bucking is expected to form a surface dipole [60].

However, understanding the mechanism for the chemically modulated work function of carbon is a great challenge though numerous models have been proposed, including the negative affinity [61,62,63], antenna effect of conducting channels [64], impurity gap states [57,65], band bending at depletion layers [66] and surface dipole formation [67]. The N-lowered $V_T$ has been explained as arising from a certain yet unknown sub-band formed above $E_F$ due to N addition. According to the impurity gap state argument, N locates at a distorted substitutional site in the host matrix with one long but weak C-N bond, which forms a deep singly occupied donor level, ~1.7 eV below the $E_C$. On the other hand, two neighboring nitrogen atoms relax away from each other due to the weak lone pair interaction, which form doubly filled states located ~1.5 eV above the $E_v$. These two mid-gap impurity levels are suggested to play dominant roles in lowering the $V_T$. Furthermore, N may create a depletion layer that causes band bending at the back contact. At sufficiently high donor concentrations, this band bending narrows the tunneling distance there, and allows emission into the diamond conduction band. However, if the N-induced mid-gap impurity levels (1.7 eV < $E_C$ and 1.5 eV > $E_V$) are dominant, the carbon co-doped with P and B should perform better than the carbon doped with N, as the P- and B-derived states (0.46 eV < $E_C$ and 0.38 eV > $E_V$) appear to be more beneficial. Boron is a shallow substitutional acceptor in diamond with a level at 0.38 eV above the valence band edge $E_V$, and phosphorus can act as a shallow donor with a level 0.46 eV below the conduction band edge $E_C$ [68]. Therefore, an atomic scale understanding of the electronic process of threshold reduction due to nitrogenation is highly desirable.

The BBB correlation mechanism could be able to solve the discrepency regarding chemical effect on the work function. The 3sp orbitals of a P atom are hard to be hybridized compared to the 2sp orbitals of O and N because the 3sp electrons are more mobile than are the 2sp electrons of N and O. The de-localized 3sp electrons determines that the P acts as a n-type donor that adds simply a DOS feature to a position 0.46 eV below the $E_C$ of a diamond [68]. That P-doping gives little reduction of the work function compared to O or N doping [57] means that the impurity gap levels narrow the band gap but contribute insignificantly to the work function reduction. Unlike P and B, O and N could expand the band gap of a semiconductor instead, through compound formation. Therefore, N and O act not simply as impurity donors or accepters in semiconductors of which no charge transport is involved. The work function of Cs and Li (~3.5 eV) is much lower than that of other metals (~5.0 eV). However, adding Cs and Li to the diamond surface is effect-less in improving the emission properties. In fact, one is unable to prevent carbide formation in the mixture of metal and carbon. In the process of carbide formation, the conducting electrons of the metal will 'flow' into the empty p-orbital of carbon, which lowers the occupied DOS of the doping metals. However, doping with both low-$\Phi$ metals, such as Li, Cs, and Ga, and N or O could form metal dipoles at the surface, which reduces the work function of the low-$\Phi$ metals even further (~1.25 eV < 3.5 eV). Therefore, co-doping low-$\Phi$ metals with O or N could be promising measures [69,70] in lowering the work function of carbon. However, it is anticipated that the production of the H-like bond at the surface due to O or N over-dosing may have detrimental effect on the work function reduction. Appropriate doping would be necessary to avoid H-like bond formation that raises the work function [31].

With the BBB correlation as origin, the impurity gap levels (lone pair) [57,65] and the surface dipole formation [67] models would be correct and complete. The lone pair impurities contribute indirectly to the work function reduction as they induce the anti-bonding dipoles. Other effects such as conducting channels [64], band bending at depletion layers [66] may play some supplementary roles in lowering the $V_T$, as these effects exist depending less on the presence of oxygen or nitrogen. New understanding may help designing and controlling work function for electron source applications.

Figure 11 Filed emission properties of nitrogentated Carbon as a function of (a) Nitrogen partial pressure, (b) substrate temperature ($P_N$ = 0.3 Pa), (c) Substrate bias, and (d) N ion implantation dose dependence of effective work function.

3.5    Band-gap expansion for photoemission

The band gap of an intrinsic semiconductor is normally within the infrared range ($E_G$ ~ 1.0 eV). After inclusion of *N*, the band-gap is widened significantly as reasoned above. For instance, the band gap of *SiN* increases from 1.1 eV to 3.5 eV with increasing *N* content [30]. The band-gap-enlargement of nitrides has been widely noted. Figure 12 shows the band-gap enlargement of (a) III-nitrides and (b) amorphous Ge- and Si-nitrides. It is noted that nitrogen incorporation into the group-III metallic solids generates a considerably large band-gap when the compound is formed. The width of the band-gap depends on the bond length [13] or the electronegativity of the corresponding element ($\eta_{Al}$ = 1.5, $\eta_{Ga}$ = 1.6, and $\eta_{In}$ = 1.7). Nitrogen expands the band-gap of the semiconductive a-Ge and a-Si from ~1.1 eV to ~ 4.0 eV [71] depending on nitrogen content. Nitrogen also widens the band-gap of amorphous carbon (*a*-$CN_x$:H) [72]. Corresponding band-gap changes of *a*-$CN_x$:H films have been observed in the He-II valence band spectra showing a recession of the leading edge of more than 0.9 eV while the optical band-gap widens from 0 to more than 1 eV.

Reynolds *et al.*[12] suggested that the green-band of ZnO and the yellow-band of GaN share some common yet unclear mechanisms. Chambouleyron et al [71] related the band-gap expansion of *a*-Ge:N and a-Si:N compounds to the substitution of Si–Si or Ge–Ge bonds by stronger Si–N or Ge–N bonds. It was suggested that, as the N content increases, the nitrogen lone-pair band develops and that the lone-pair band dominates the valence-band maximum as the stoichiometry is reached. The largest optical band-gap is obtained for the stoichiometric compound. On the contrary, for smaller N content, Si–Si or Ge–Ge bonds dominate the valence band maximum.

For amorphous semiconductors, it is generally accepted that the transition of carriers is between the conduction-band tail and valence-band tail states. Luminescence spectra [30] of the a-Si:H showed that the n-type (phosphorous) doping shifts the luminescence peak of the a-Si:H from 1.1 eV to 0.81 eV, and the p-type (boron) doping shifts the peak to 0.91 eV. This can be easily understood in terms of impurity levels. The shallow n-donor levels and the deeper p-acceptor levels are located within the initial band-gap (1.1 eV width) near to the band tails, which should narrow the gap, as observed. However, the luminescence peak of the a-Si:N:H compound moves to higher energy with increasing nitrogen concentration [73]. The broadened band-gap through nitridation could not be explained in terms of the traditional donor effect, though the nitrogen addition is always believed as n-type doping. Clearly, the band-gaps of metal nitride, III-nitride, and IV-nitride are enlarged by the same hole-production mechanism proposed in the BBB correlation mechanism. The change of bond nature and bond length has an effect on the crystal field, and consequently, the width of the band-gap; charge transport in the reaction re-populates with valence electrons of the host materials.

Figure 12 (a) N-concentration dependence of the optical band-gap ($E_{Tauc}$) of *a*-Ge:N:H [74,75] and *a*-Si:N:H [76] thin films.

*IV*    *Summary*

Consistent understanding of the unusual performance of a nitride has been developed from the perspective of bond-band-barrier correlation. It has been clear that the *N*-enhanced magnetization, the blue shift in nitride light emission, the *N*-lowered threshold of cold cathode in diamond, wear and corrosion resistance and super elastic nitrides all arise from the nitride tetrahedron bond formation with involvement of bonding, nonbonding, hole production and antibond dipole production. Differing from the density functional calculations, the current model describes directly the electronic configuration in the bonding process and its derivatives on the valence *DOS* and physical properties. The concept of minimal total energy being a criterion for optimal solution in calculations is not needed in the current *BBB* correlation. Further extension of this correlation and the concept of lone pair nobonding, antibonding dipoles and hole states may provide guideline for the controllable modification of existing materials and in the pursuit of new functional properties. The current interpretation may be helpful in thinking about the nitride in a bond forming way. This will stimulate more interesting topics, such as quantification of the bond geometry and determination of bonding energy for specific systems, towards controlling bond-and-band forming towards designer controlled materials design and fabrication.

Acknowledgment- WTZ acknowledges founding support from the Teaching and Research Award Program for OYTHEI of MOE and the special foundation for PhD program in High

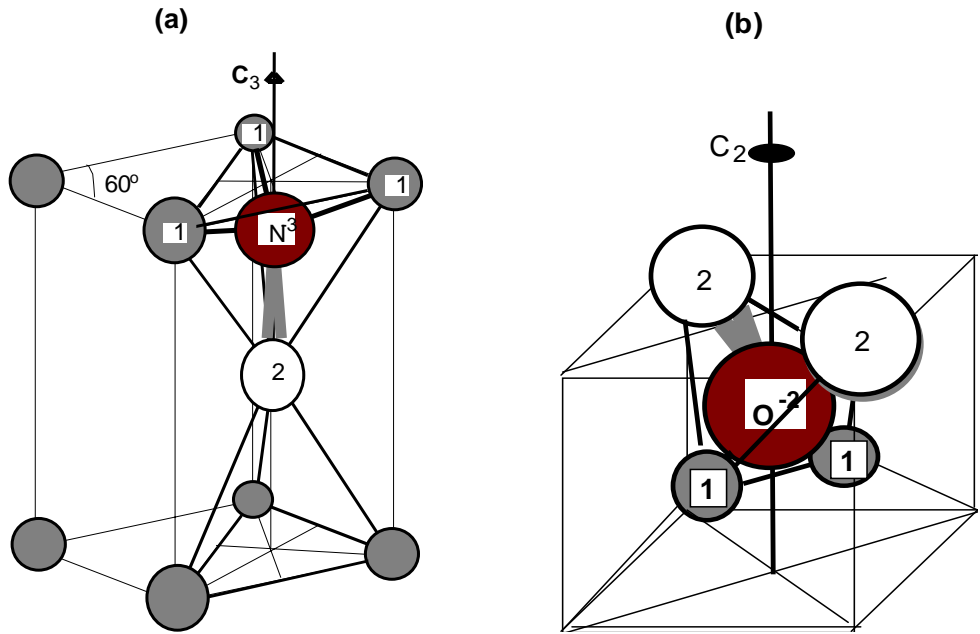

Fig-01

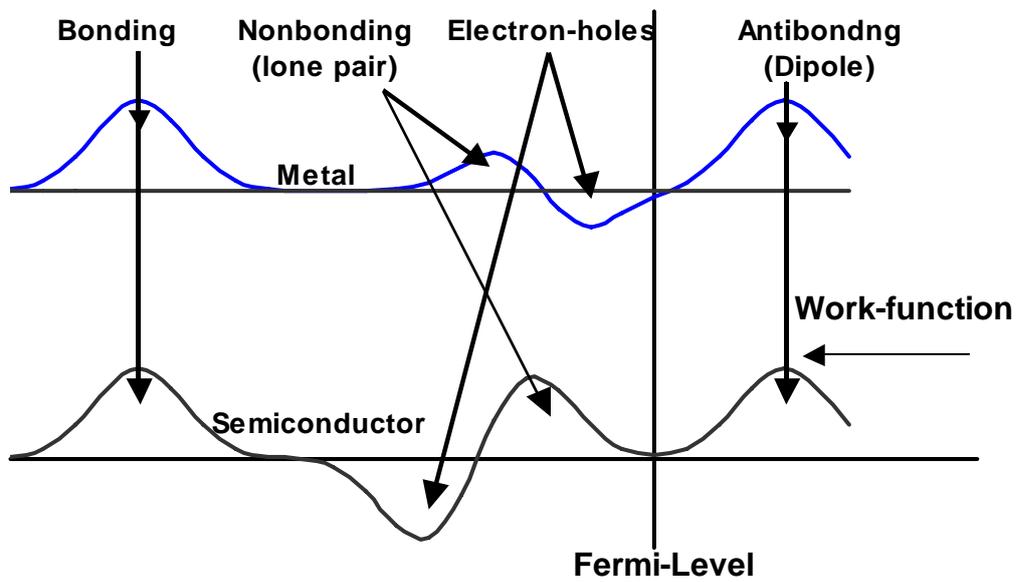

Fig-02

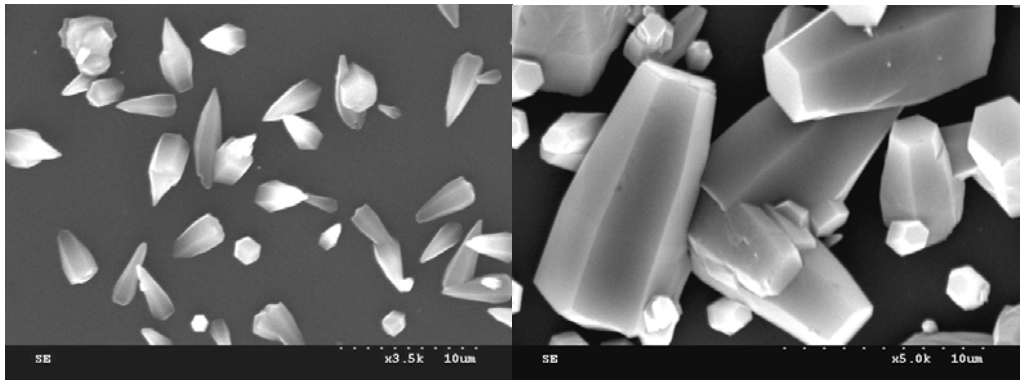

(a)  (b)

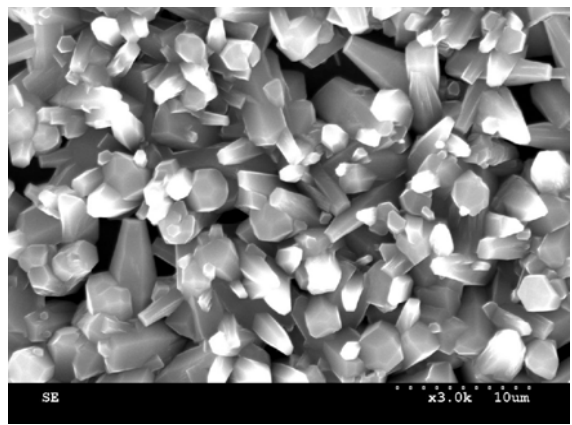

(c)

Fig-03

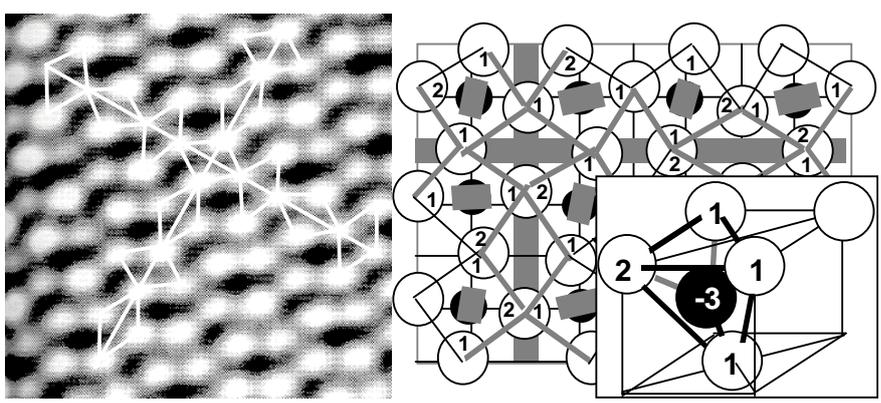

Fig-04

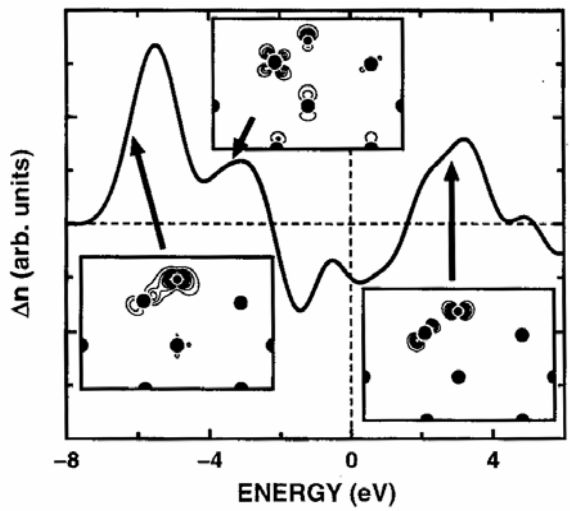

Fig-05

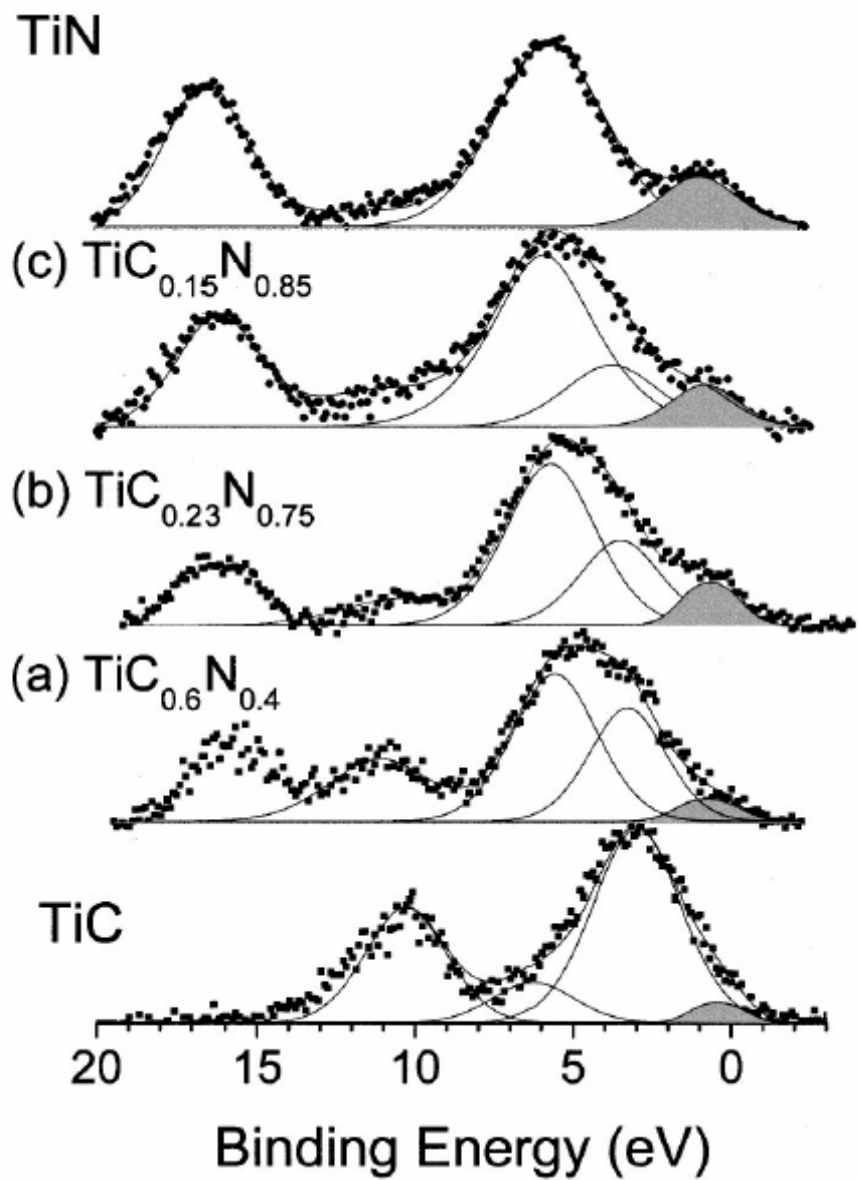

Fig-06

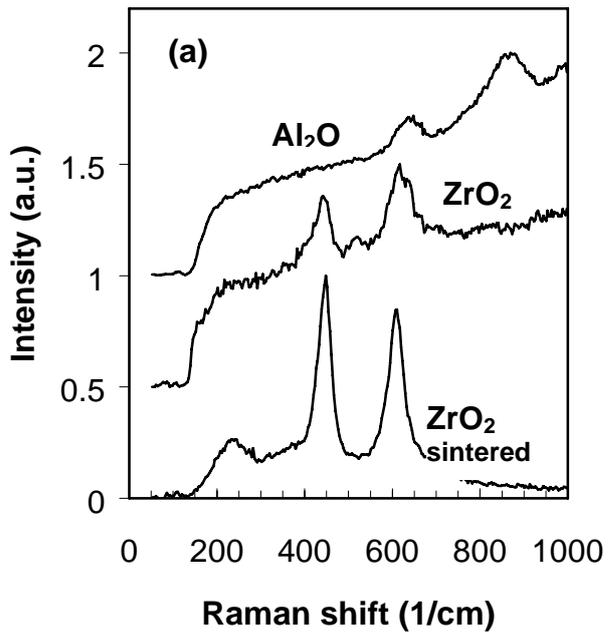

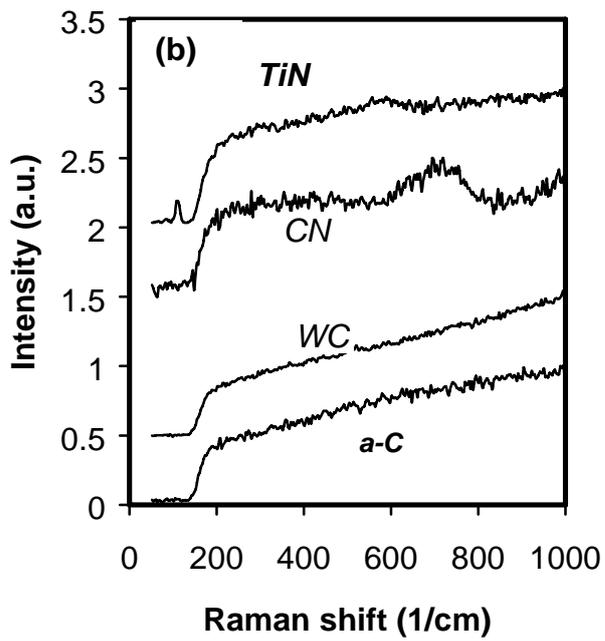

Fig-07

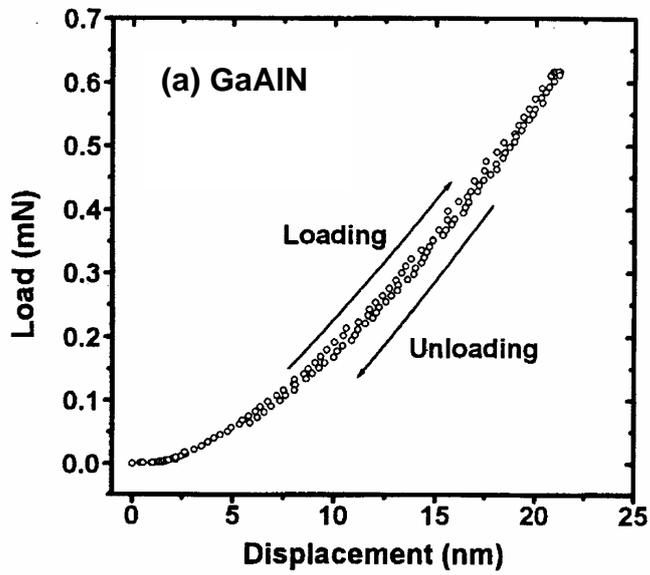

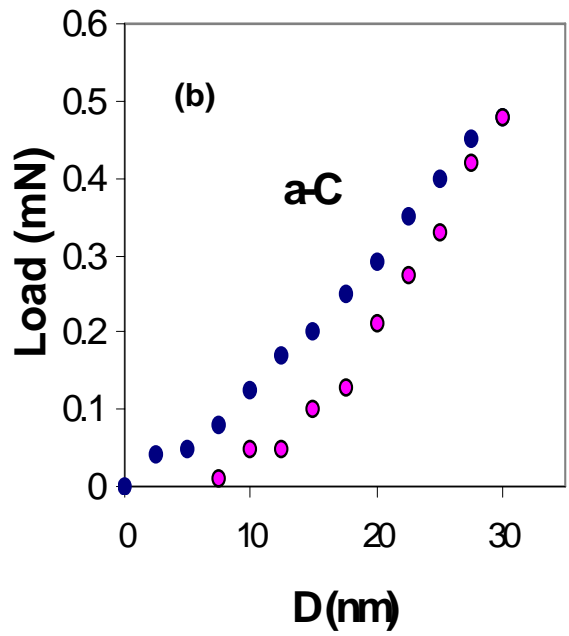

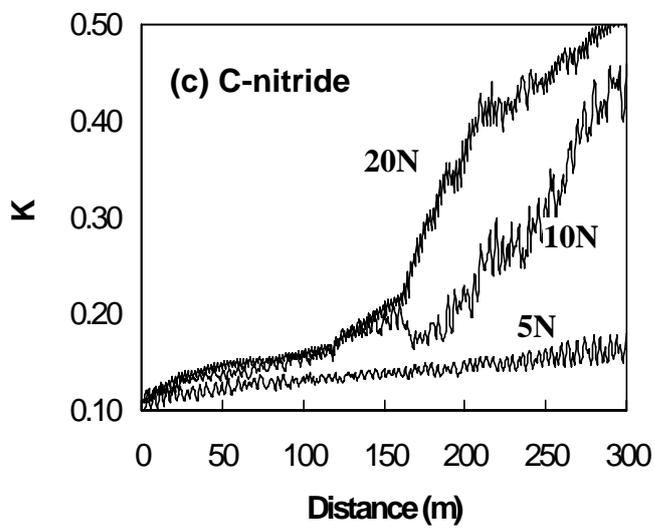

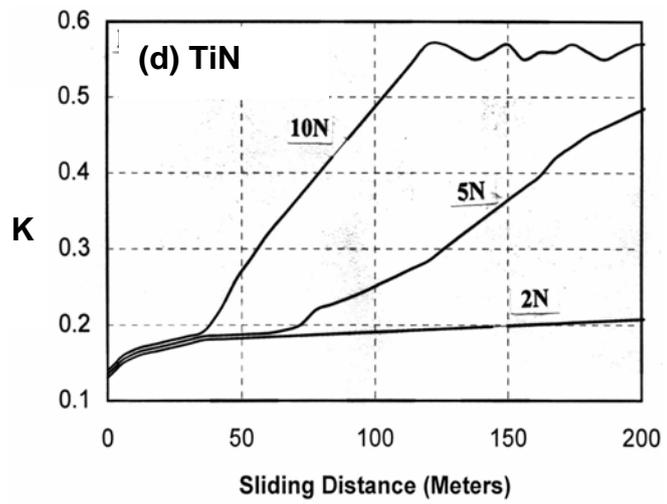

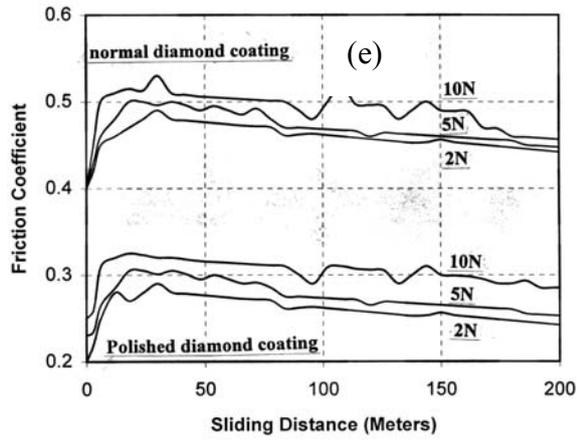

Fig-08

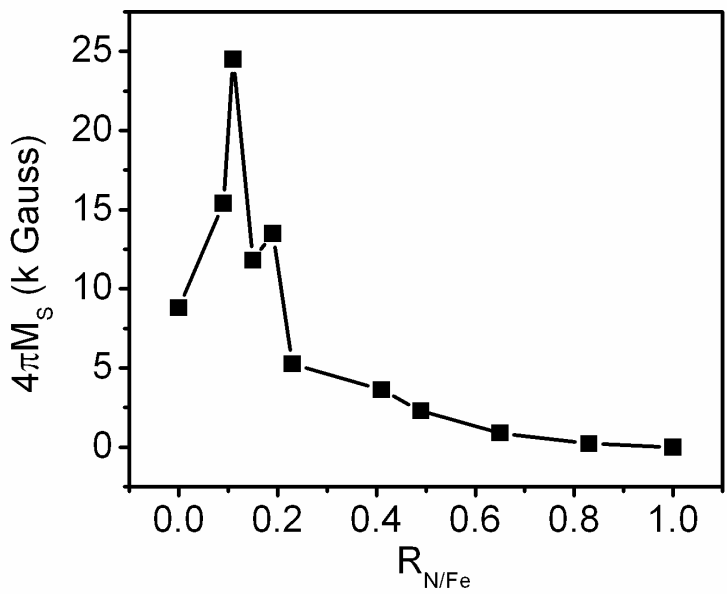

Fig-9

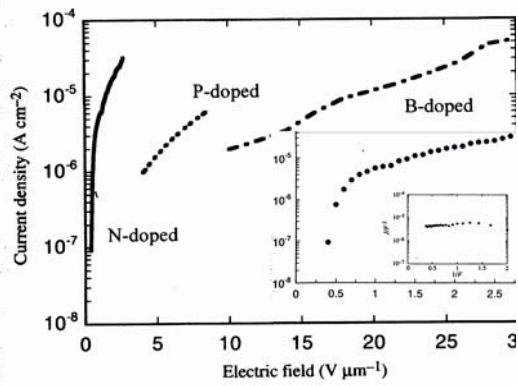

Fig-10

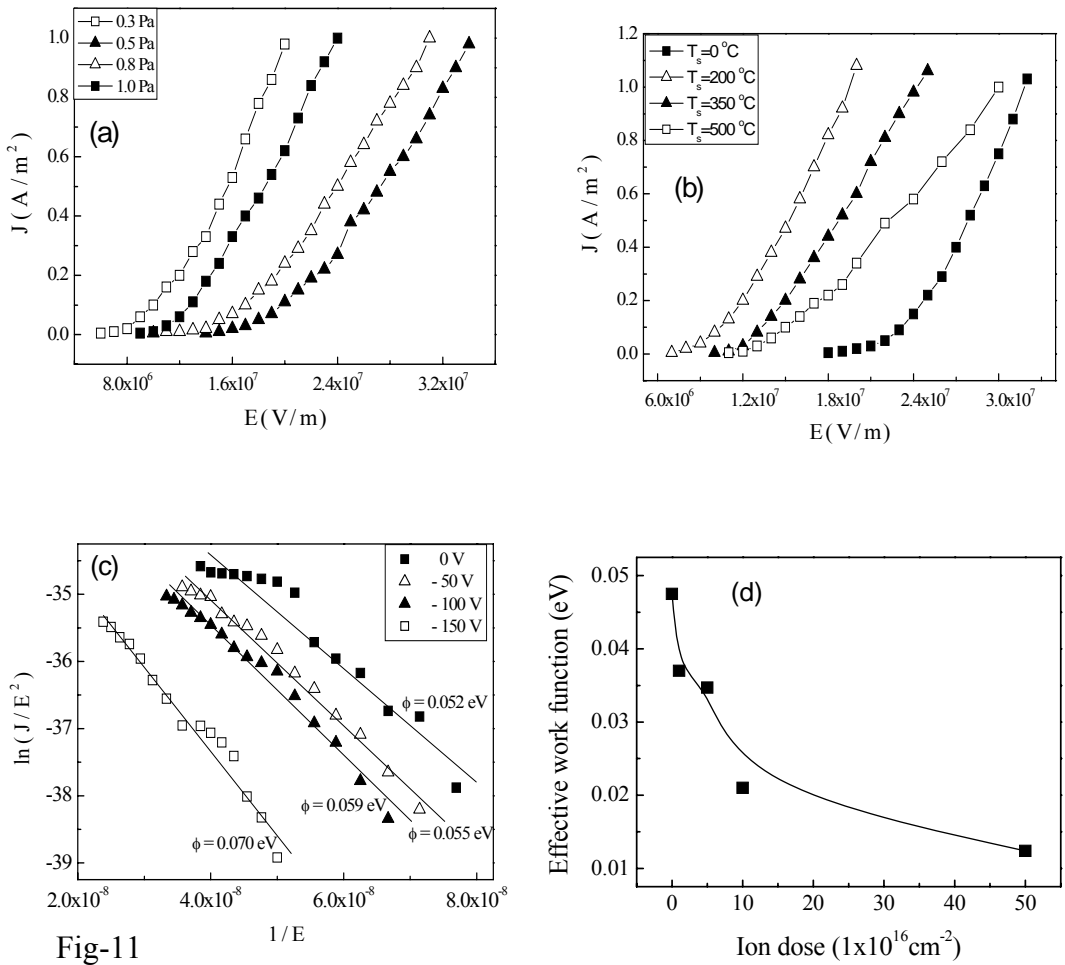

Fig-11

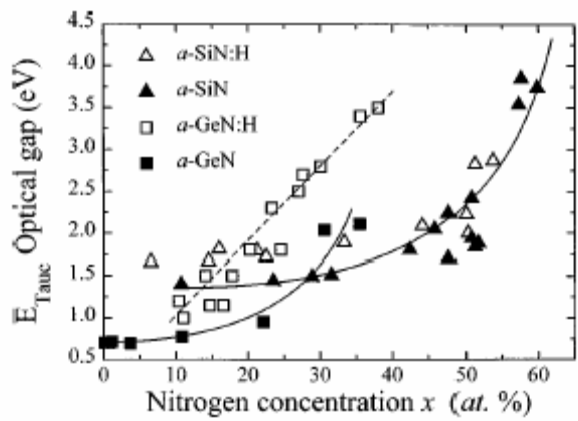

Fig-12